\documentclass[aip,reprint,graphicx]{revtex4-1}
\usepackage{mathtools}
\usepackage{graphicx}
\usepackage{gensymb}
\usepackage[export]{adjustbox}
\usepackage{amsmath}
\usepackage{float}
\usepackage{hyperref}
\usepackage{hypcap}
\usepackage{flushend}
\draft
\begin{document}
\title{Theoretical Spectroscopic Investigation of Hydrogen Bonding and Hydrophobicity} 

\author{Kambham Devendra Reddy}
\affiliation{Department of Chemistry, Indian Institute of Technology Tirupati, Tirupati 517506, India}
\author{Rajib Biswas}
\email[Author to whom correspondence should be addressed. Electronic mail: ]{rajib@iittp.ac.in}
\affiliation{Department of Chemistry, Indian Institute of Technology Tirupati, Tirupati 517506, India}

\date{\today}
\begin{abstract}
Hydrophobic solutes significantly alters water hydrogen bond network. The local alteration of solvation structures get reflected in the vibrational spectroscopic signal. Although it is possible to detect this microscopic features by modern infrared spectroscopy, however, bulk phase spectra often comes with formidable challenge of establishing the connection among the experimental spectra to molecular structures. Theoretical spectroscopy can serve as more powerful tool even where spectroscopic data cannot provide microscopic picture. In the present work, we build a theoretical spectroscopic map based on mixed quantum-classical molecular simulation approach using methane in water system. The single oscillator level O-H stretch frequency is well correlated with a collective variable solvation energy. We construct the spectroscopic maps for fundamental transition frequencies and also the transition dipoles. A bimodal frequency distribution with a blue shifted population of transition frequency illustrates presence of gas like water molecules in the hydration shell of methane. This observation is further complemented by a shell-wise decomposition of the O-H stretch frequencies. We observe a significant increase in ordering of the first solvation water except the water molecules, which are directly facing the methane molecule. This is manifested in redshift of the observed transition frequencies. Temperature dependent simulations depict that the water molecules facing to the methane molecule behave similar to the high temperature water and the rest of the first shell water molecules behave more like cold water.
\end{abstract}
\maketitle 

\section{Introduction}
The incessant evolution of hydrogen bond network makes water as one of the most interesting liquid.\cite{Bagchi2011,Ball2015} In spite of its smaller size, water molecules are capable of behaving hydrogen bond donor and acceptor simultaneously. As a result a gigantic hydrogen bond network is abundant in bulk water. The ultrafast evolution of this massive hydrogen bond network leads to many unique features of water.\cite{Nandi2000,Bagchi2005,Bellissent-Funel2016} The presence of external solutes perturbs the length scale and the time scale of evolution of the hydrogen bond network that makes the systems even more fascinating.\cite{Ikeguchi1998,Kropman2001,Raschke2005,Bakulin2011,Grdadolnik2017}
\par
The hydrophobic effect is the manifestation of the interaction of non-polar moieties with water. The hydrophobic effect is typically inferred at two levels. First, the interaction between a non-polar molecule and the surrounding water molecules is called \textit{hydrophobic hydration}. Second is known as \textit{hydrophobic interaction} or \textit{pair hydrophobicity} which describes the interaction between two non-polar molecules in water as a function of their separation distance. The hydrophobic effect shows pivotal role in biological processes such as protein folding, formation of cell membrane, formation of vesicles and lipid bilayer, assembly of proteins into functional complexes, and many more.\cite{Tanford1973,Srinivas2002,Srinivas2003,Mason2004,Chandler2005,Berne2009,Hazra2014} The significant role in Biology and Chemistry makes the study of hydrophobic effect a subject of major interest till date. In spite of several earlier conclusive experimental and theoretical studies, the microscopic origin and the length scale of hydrophobic interaction are still not understood completely and yet remain an active area of research.
\par
The presence of many body interactions could give rise to different structural arrangement surrounding the hydrophobes.
\cite{Bakulin2011,Grdadolnik2017,Chandler2005,Wiggins1997,Lum1999,Huang2000} It is considered that the presence of hydrophobic solutes modifies the shape of the solvation shell into tiny \textit{icebergs}.
\cite{Frank1945,Kauzmann1959} This type of solvent arrangement significantly alters the hydrogen bond strength in the close vicinity of the hydrophobic moieties.\cite{Raschke2005,Bakulin2011,Grdadolnik2017,Koh2000,Scatena2001,Davis2012,Montagna2012} It has also been argued that for sufficiently weak solute-water attraction, a large smooth hydrophobic surface might be enclosed by a microscopically thin film of water vapor.\cite{Stillinger1973} The recent finding also shows that the hydrophobic solutes reinforce the water hydrogen bonds of solvation shell water molecules.\cite{Grdadolnik2017,Koh2000} 
As the vibrational frequencies of water are highly sensitive to the local microscopic solvation configuration, infrared (IR) spectroscopy is a tempting method for exploration of these systems.\cite{Fecko2003,Biswas2012} IR spectroscopy of O-H stretching mode is usually employed as the most trustworthy and sensitive method for estimating relative strengths of hydrogen bonds.\cite{Davis2012,Fecko2003,Hecht1992,Hecht1993,Sharp2001,Auer2007,Laage2009,Stiopkin2011,Biswas2013,Biswas2016}
\par
In spite of its extensive use, meaningful interpretation of IR spectroscopic data face the formidable challenge of establishing the connection among the experimental spectra to molecular structures in the bulk phase. Experimental spectra represent a superposition of different transient solvation structures, hence illustrate the macroscopic response. Besides, the existence of strong anharmonic couplings leads to further delocalization through different vibrational modes mixing. On the contrary, the microscopic resolution of computer simulation aided spectroscopy modeling empowers us to investigate these systems at the molecular level.\cite{Auer2007,Biswas2013,Biswas2016,Corcelli2004,Corcelli2005,Roberts2009,Biswas2017,Samanta2018} Therefore, the theoretical spectroscopy offers a window to decipher the experimental data in rather quantitative fashion.\par
In spite of several earlier computational studies, the structure spectrum correlation of aqueous hydrophobic system needs further quantification using appropriate spectroscopic modeling. As in bulk water, vibrational modes of O-H oscillators are significantly coupled and give rise to very broad response, which makes it challenging to differentiate contribution from diverse transient configurations. To reduce the complexity of the problem, isotope dilution strategies are often used.\cite{Auer2007,Biswas2016,Corcelli2005,Samanta2018} The isotope dilute system eventually allows us to study the system in lower dimensionality by isolating a local oscillator. Thus, a small percentage of H$_2$O in a fully deuterated solution can provide isolated O-H oscillators, which can function as a local probe to different solvation structures.\par 
In a number of earlier works, mixed quantum-classical (MQC) models have been extensively utilized to explain the O-H stretching vibrations in various isotope dilute aqueous systems.\cite{Auer2007,Biswas2016,Corcelli2004,Corcelli2005,Roberts2009,Samanta2018} Usually, these models identify a collective coordinate from classical molecular dynamics (MD) trajectories to depict the influence of the solvent on the quantum mechanical spectroscopic coordinate. The electronic structure calculation coupled with MD simulation is the fundamental basis of the MQC approach. The MQC models are widely used to construct the required spectroscopic maps, which can essentially be used to generate the trajectories of time-dependent transition frequencies and dipole moments.\par
In this work, we investigate the effect of hydrophobicity in water structure by employing computer simulation aided spectroscopy modeling which employs the microscopic environmental sensitivity of vibrational frequencies. The rest of this article is structured as follows: In section II, we will elaborate the spectroscopic modeling and simulation details, data analysis and major findings are discussed in section III and conclusions are given in section IV.
\section{methods}
\subsection{Classical molecular dynamics}
We perform molecular dynamics simulation using GROMACS version 2019.1.\cite{Abraham2015} The system consists of a cubic box of 255 SPC/E\cite{Berendsen1987} water molecules and 1 OPLS-AA methane molecule.\cite{Jorgensen1996} For bulk water, we take 256 water molecules. Periodic boundary conditions are applied in all three directions.\cite{Frenkel2002} We perform energy minimization of the systems using steepest decent method. Thereafter, the systems are equilibrated in \textit{NPT} ensemble for 1ns. Finally, the data acquisitions are done in \textit{NVT} ensemble over 5 ns long trajectory. We use Berendsen thermostat\cite{Berendsen1984} with a relaxation time of 0.1 ps and Parinello-Rahman barostat\cite{Parrinello1981,Nose1983} with a relaxation time of 1.0 ps for maintaining constant temperature at 300 K and pressure at 1 bar, respectively. We use a 9 \text{\AA} cutoff radius for neighbor searching and non-bonded interactions and all the bonds are kept fixed using LINCS.\cite{Hess1997} The long-ranged electrostatic interactions are calculated using Particle Mesh Ewald (PME) with FFT grid spacing of 1.6 \AA.\cite{Darden1993}

\subsection{Mixed quantum mechanical calculation}
We use mixed quantum classical approach to build the single oscillator O-H stretch spectroscopic modeling of methane in water. This technique is based on semi-empirical cluster-based mapping method, which has been widely used in simulating the O-H vibrational spectroscopy of isotope diluted aqueous systems.\cite{Biswas2016,Corcelli2004,Corcelli2005,Roberts2009} We utilize electronic structure calculations on instantaneous small frozen clusters extracted from the classical MD trajectory to construct the spectroscopic map. In this, we correlate the quantum mechanical O-H transition frequency and transition dipole moment onto a collective coordinate which can easily be calculated from the classical trajectory. Subsequently, by using this classical analogue of the collective variable, we generate the transition frequency and transition dipole trajectories, which can further be utilized to calculate linear and nonlinear IR spectra from respective time-domain response functions.\par
From the classical trajectory, we extract small clusters by identifying a central H atom that belongs to the water molecule closest to the methane molecule and include any molecules having its oxygen within a 7.0 \text{\AA} radius. We use such 165 methane-water clusters. For bulk water response, we choose 130 small clusters extracted in the similar fashion from bulk water trajectory. These clusters contain average of $\sim44$ water molecules, which is adequate to approximately resemble the bulk like environment around the central O-H bond, and slight fluctuation in the number of water molecules in a cluster does not affect our findings. Furthermore, these clusters are chosen in such a way so that we can sample all relevant configurations. A more detail description concerning the cluster selection is provided in the supplementary information.\par
We get the O-H stretch response for each selected cluster, from the one-dimensional adiabatic potential energy surface (PES). We construct the quantum mechanical PES, by stretching the central O-H bond from $r_{\mathrm{OH}} = 0.7$ to 1.6 \text{\AA} with a grid spacing of 0.08 \text{\AA}, while keeping the remaining degrees of freedom frozen. For getting the single point energy, we perform DFT calculations employing B3LYP hybrid functional\cite{Becke1993,Lee1988,Vosko1980,Stephens2002} and 6-311++G(d,p) basis set in Gaussian 16 package.\cite{Frisch2016} The choice of B3LYP functional is inspired by the earlier work of Skinner and co-workers, which demonstrated that potential energy obtained using B3LYP functional along the O-H stretch coordinate is in promising agreement with the coupled cluster CCSD(T) prediction.\cite{Gruenbaum2013} Although the classical simulations are executed with isotopically pure systems, we consider that all hydrogens but the proton of the selected O-H bond are deuterons. We obtain the eigenstates, $|n \rangle $, and eigenvalues, $E_n$, of PES $U(r_{\mathrm{OH}})$ by solving the one dimensional Schrodinger equation using the discrete variable representation (DVR)\cite{Colbert1992,Groenenboom1993} with a grid spacing of 0.01 \text{\AA} and reduced mass of the O-H vibration of the HOD molecule (0.954426 amu).\cite{Roberts2009} Then from the energy eigenvalues, we obtain the transition frequencies $\omega_{nm} = (E_n-E_m)/ \hbar$ between different vibrational states.\par
We calculate the transition dipole moments in a manner similar to Corcelli and Skinner.\cite{Corcelli2005} The transition dipole matrix elements for the transition between vibrational states n and m is expressed as:
\begin{equation}
   \vec{\mu}_{nm} = \langle n | \vec{\mu} |m \rangle
\end{equation}
After expanding the dipole operator about the minimum of the O-H stretching potential ($r_{eq}$ to the first order term in $r_\mathrm{{OH}}$, one can approximate $\vec{\mu}_{nm}$ as
\begin{equation}
    \vec{\mu}_{nm} = \langle n | \vec{\mu}_0 + r_{\mathrm{OH}}\left(\frac{d\vec{\mu}}{dr_{\mathrm{OH}}}\right)_{r_{\mathrm{OH}}=r_{eq}} |m \rangle \approx \mu^\prime r_{nm} \hat{\textbf{u}}
\end{equation}
where $r_{nm} = \langle n | r_{\mathrm{OH}} |m \rangle$ are the matrix elements obtained from the eigenstates of the DVR calculation, $\vec{\mu}_{nm}$ can be divided into a magnitude $\mu$ and a direction $\hat{\textbf{u}}$, which we consider as lying along the O-H bond axis and $\mu^\prime$ is the dipole moment derivative. We acquire $\mu^\prime$ for each chosen cluster by calculating $\vec{\mu}_{nm} \cdot \hat{\textbf{u}}$ at five $r_{\mathrm{OH}}$  displacements separated by 0.01 \text{\AA} about $r_{eq}$, and then numerically differentiate with respect to $r_{\mathrm{OH}}$ . Finally, we construct correlation maps for $\omega_{nm}$ and $\mu_{nm}$ against a collective coordinate.\par
Although it was reported earlier that the projected electric field on the O-H stretch could serve as a potential collective variable, which can describe the spectroscopy of isotope diluted aqueous system.\cite{Auer2007,Corcelli2005} However, we find it is not the case for the water methane system (see supplementary material). Therefore, we use an alternative collective variable as used in the case of aqueous hydroxide and aqueous proton systems.\cite{Roberts2009,Biswas2016} We use solvation energy as the collective variable defined as the difference in the potential energy at two reference points along $r_{\mathrm{OH}}$ .
\begin{equation}
    \Delta E_{\mathrm{DFT}}=U(r_f)-U(r_i)
\end{equation}
As the potential energy is highly sensitive to the local environment, of the solvation energy accurately captures the alteration in the PES as a function of the hydrogen bond strength. After rigorous tweaking, we find that the $r_i = 1.0 $ \text{\AA} and $ r_f = 1.4 $ \text{\AA} combination provides a better correlation with the spectroscopic variables.

\section{Results and Discussion}
\subsection{Spectroscopy maps}
We have mentioned earlier that hydrophobic solutes alter the water hydrogen bond network and thus change the local electronic environments of O-H oscillators. The intense local environmental sensitivity makes vibrational spectroscopy of O-H stretching mode $(\nu_{OH})$ a most reliable and sensitive approach for investigating relative strengths of H-bonds. The alteration of H-bond strength will result in shifting of observed frequency for $(\nu_{OH})$ mode. Thus, enhancements of the H-bond strength will eventually be reflected in frequency redshift of the $(\nu_{OH})$ mode when compared the spectra of bulk water with those of the water molecules perturbed by hydrophobic solute. Furthermore, structural order can also be investigated by examining the spectral line width. As more structural ordering will have less variations in the local structures and will eventually results into a narrower line width.\par
\begin{figure}
	\includegraphics[width=1.0\linewidth]{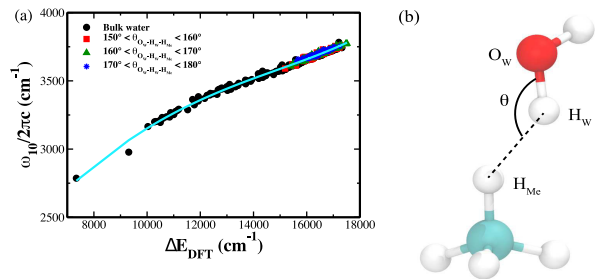}
	\caption{\label{fig1}(a) Correlation between fundamental O-H transition frequency and solvation coordinate obtained from DFT calculation  using clusters taken from the classical MD. Black color filled circles represents the bulk water data, whereas all other colored filled symbols represent the data for methane-water cluster. The cyan solid line represents the fourth order polynomial fit. (b) Definition of the angle between methane hydrogen, water hydrogen and water oxygen.}
\end{figure}
The general observation is that the observed one-dimensional O-H stretch potential energy surfaces becomes less anharmonic in presence of methane, which results in O-H stretch frequency in the high frequency range of the bulk-water response. The correlation of the fundamental transition frequency and solvation energy obtained from the quantum mechanical calculation is presented in Figure \ref{fig1}. The O-H stretch frequencies are monotonic and highly correlated (with correlation coefficient 0.9964) through a nonlinear relationship with the collective solvation coordinate within the 1000 cm-1 range spanned by the different cluster configurations. Although, we select all the instantaneous configurations without any bias, we have further investigated the effect of orientation of the tagged O-H bond with the methane molecule. Note that, this orientational dependencies are investigated only for the O-H oscillators, which are directly facing to the methane molecule. This was achieved by sampling the configurations having angle $\angle \mathrm {O_WH_WH_{Me}}$ (as defined in Figure \ref{fig1}(b)) within a particular range. We find there exist minimal or almost no effect of the orientation of the O-H oscillator with respect to the methane moiety on the spectroscopy map. We use a $4^{th}$ order polynomial fitting function to get the empirical relations of the transition frequencies with solvation coordinate (given in Table \ref{table1}). We furthermore explore the spectroscopic maps for other transition frequencies and these are shown in the supplementary material (Figure \ref{figs3}).\par
\begin{figure}[b]
	\includegraphics[width=0.8\linewidth]{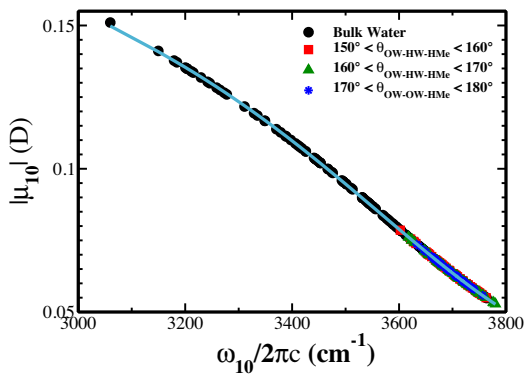}
	\caption{\label{fig2}Correlation between fundamental O-H stretching frequency ($\omega_{10}$) and transition dipole matrix elements. Cyan line represents the fourth order polynomial fit.}
\end{figure}
\begin{table*}[]
\centering
\caption{Empirical relation between the transition frequency and the solvation energy obtained by fitting the quantum mechanical data as shown in Figure \ref{fig1}. The fitting function is $\omega_{10}=a_0+a_1 \Delta E_{\mathrm{DFT}}+a_2\Delta E_{\mathrm{DFT}}^2+a_3\Delta E_{\mathrm{DFT}}^3+a_4\Delta E_{\mathrm{DFT}}^4$. The parameters for other transition frequencies are given in supplementary information (Table S2).}
\begin{tabular}{llllll}
\hline
$a_0$    & $a_1$   & $a_2$        & $a_3$          & $a_4$         & Correlation Coefficient \\
\hline
\hline
1090.37 & 0.2636 & $4.7833\times 10^{-7}$ & $-9.0792\times10^{-10}$ & $3.0082\times10^{-14}$ & 0.9964\\
\hline
\end{tabular}
\label{table1}
\end{table*}
Subsequently, we construct the empirical maps for transition dipole moments. We follow the strategy as explained earlier. It was shown earlier that empirical relation can be established between the dipole moment derivative $\mu^\prime$ and the electric field projected along $r_{\mathrm{OH}}$  in the $\mathrm{HOD/D_2O}$ system or the solvation coordinate for the aqueous hydronium and hydroxide systems. In the present work, we find the best correlations between $\mu^\prime$ to $\Delta E_{\mathrm{DFT}}$ and $\mu_{nm}$ to $\omega_{nm}$. The correlation between the fundamental transition dipole moment $\mu_{10}$ and the fundamental transition frequency $\omega_{10}$ is shown in Figure \ref{fig2}. The fundamental transition dipole moments are monotonic and highly correlated (with correlation coefficient 0.9999) through a nonlinear relationship with the fundamental transition frequency (Figure \ref{fig2}). We use a $4^{th}$ order polynomial fitting function to get the empirical relation between fundamental transition dipole moment and fundamental transition frequency (fitting parameters are shown in Table \ref{table2}).\par
\begin{figure}
\includegraphics[width=0.8\linewidth]{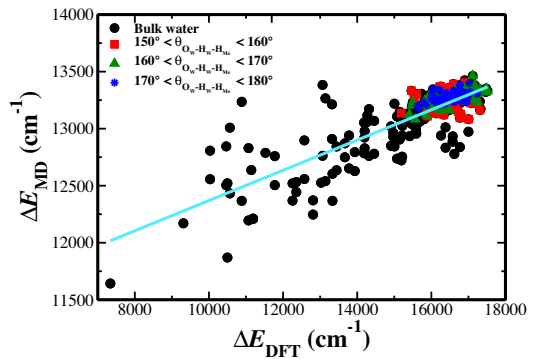}
\caption{\label{fig3}Correlation between solvation energy obtained from DFT calculation and MD calculations. Black color filled circles represents the bulk water data, whereas rests represent the data for methane-water cluster. The cyan solid line represents the linear fit.}
\end{figure}

\begin{table*}
\centering
\caption{Empirical relationships of the fundamental O-H stretching frequency with transition dipole matrix element using $\mu_{nm}=a_0+a_1 \omega_{nm}+a_2\omega_{nm}^2+a_3\omega_{nm}^3+a_4\omega_{nm}^4$.}
\begin{tabular}{llllll}
\hline
$a_0$    & $a_1$   & $a_2$        & $a_3$          & $a_4$         & Correlation Coefficient \\
\hline
\hline
21.2154	& -0.0258	& $1.1895\times10^{-5}$ &	$-2.4285\times10^{-9}$	& $1.8459\times10^{-13}$	& 0.9999\\
\hline
\end{tabular}
\label{table2}
\end{table*}
Since the quantum mechanical calculation of the solvation energy at each time step for a tagged O-H bond is computationally expensive; this necessitates an appropriate quantity that will be adequate for estimating the transition frequencies from the classical MD trajectory. To overcome this, we construct a correlation map between the quantum mechanical solvation coordinate i.e. $\Delta E_{\mathrm{DFT}}$ and that calculated from the classical simulation force field $\Delta E_{\mathrm{MD}}$ (Figure \ref{fig3}). The classical solvation coordinate $\Delta E_{\mathrm{MD}}$ is also computed in similar fashion by using $\Delta E_{\mathrm{MD}} = U_{\mathrm{MD}}(r_f)-U_{\mathrm{MD}}(r_i)$. Although the classical forcefield is not designed to compute the correct full PES for O-H stretch, our calculation nevertheless depicts that these two variables are linearly correlated with a correlation coefficient of 0.83 (Figure \ref{fig3}). The fitting parameters are represented in Table \ref{table3}.

\begin{table}[]
\centering
\caption{Linear empirical relation between solvation energy obtained from DFT calculation and MD calculations. We use $\Delta E_{\mathrm{MD}}=a_0+a_1 \Delta E_{\mathrm{DFT}}$ fitting function and the parameters and correlation coefficient are shown below.}
\begin{tabular}{lll}
\hline
$a_0$   & $a_1$    & Correlation Coefficient \\
\hline
\hline
11044.4	& 0.1326 &	0.83\\
\hline
\end{tabular}
\label{table3}
\end{table}

\subsection{Structure-spectrum correlations}
\begin{figure*}[t!]
\includegraphics[width=.9\linewidth]{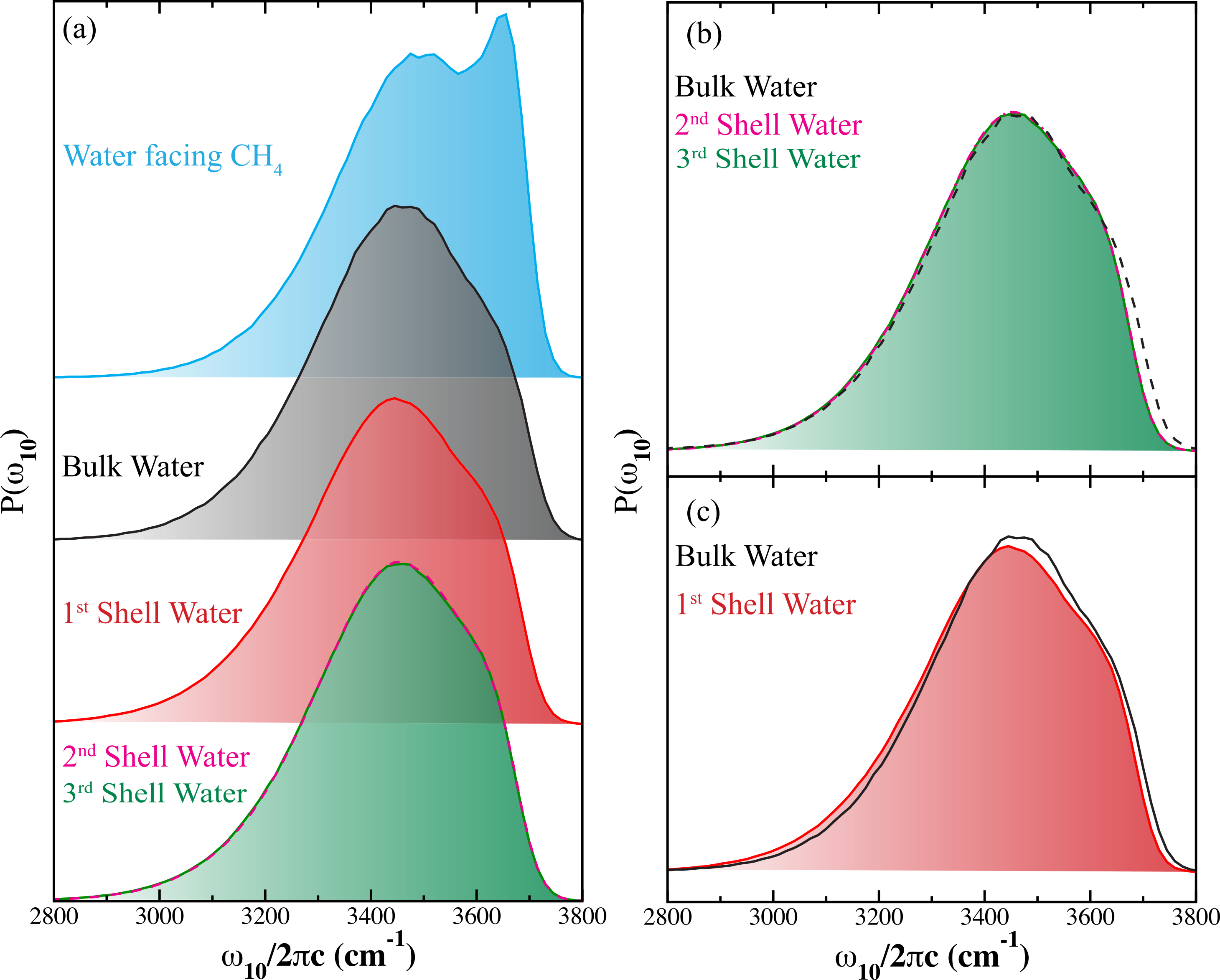}
\caption{\label{fig4}(a) Comparison of stretching frequency  of bulk water (black), closest water to methane (cyan), first solvation shell (red), second solvation shell (magenta) and third solvation shell (green). (b) Comparison of stretching frequency  of bulk water, second and third solvation shell water of methane-water system. (c) Comparison of stretching frequency  of bulk water and first solvation shell water of methane-water system.}
\end{figure*}

 To investigate the microscopic origin of different observed frequencies, we explore the structure-spectrum relationships. This was achieved by analyzing the static frequency distributions for distinct local environments sampled by a tagged O-H bond. We examine the effect of hydrophobicity on the transition frequency of O-H stretch by investigating the water molecule nearest to the methane moiety. Furthermore, we select the O-H bond of that selected water, which is pointing towards the methane. In Figure \ref{fig4}, we represent the static histogram distribution of fundamental transition frequency for bulk water and methane-water systems. The methane-water response is further decomposed into three sub-ensembles: $1^{st}$ solvation shell, $2^{nd}$ solvation shell and $3^{rd}$ solvation shell. The criteria for solvation shell decompositions are shown in supplementary information. It is evident from the Figure \ref{fig4} that the frequency distributions of closest water that are facing towards methane moiety and $1^{st}$ solvation shell water are significantly different than that of bulk water. The response of water molecules, which are directly facing the methane moiety, shows a two-ensemble blueshifted distribution (Figure \ref{fig4}(a)). Although the observed blueshift is apparent from the fact that the water molecules that are directly facing the methane moiety are lacking in number of hydrogen bonds; the two-ensemble picture needs further exploration. In case of $1^{st}$ solvation shell water, the distribution is redshifted with respect to the bulk water distribution. We indeed find that the distribution maximum shifted from bulk water response i.e. $\sim 3460 cm^{-1}$ to $\sim 3445 cm^{-1}$ in case of $1^{st}$ shell water (Figure \ref{fig4}(c)). This observation is similar to the earlier findings that the water in the first solvation shell of methane has enhanced structural ordering.\cite{Raschke2005,Grdadolnik2017,Koh2000} However, the water in the $2^{nd}$ and $3^{rd}$ solvation shell of methane almost behave like bulk water (Figure \ref{fig4}(a) and 5(b)).
 \par
 
\begin{figure}
\includegraphics[width=.8\linewidth]{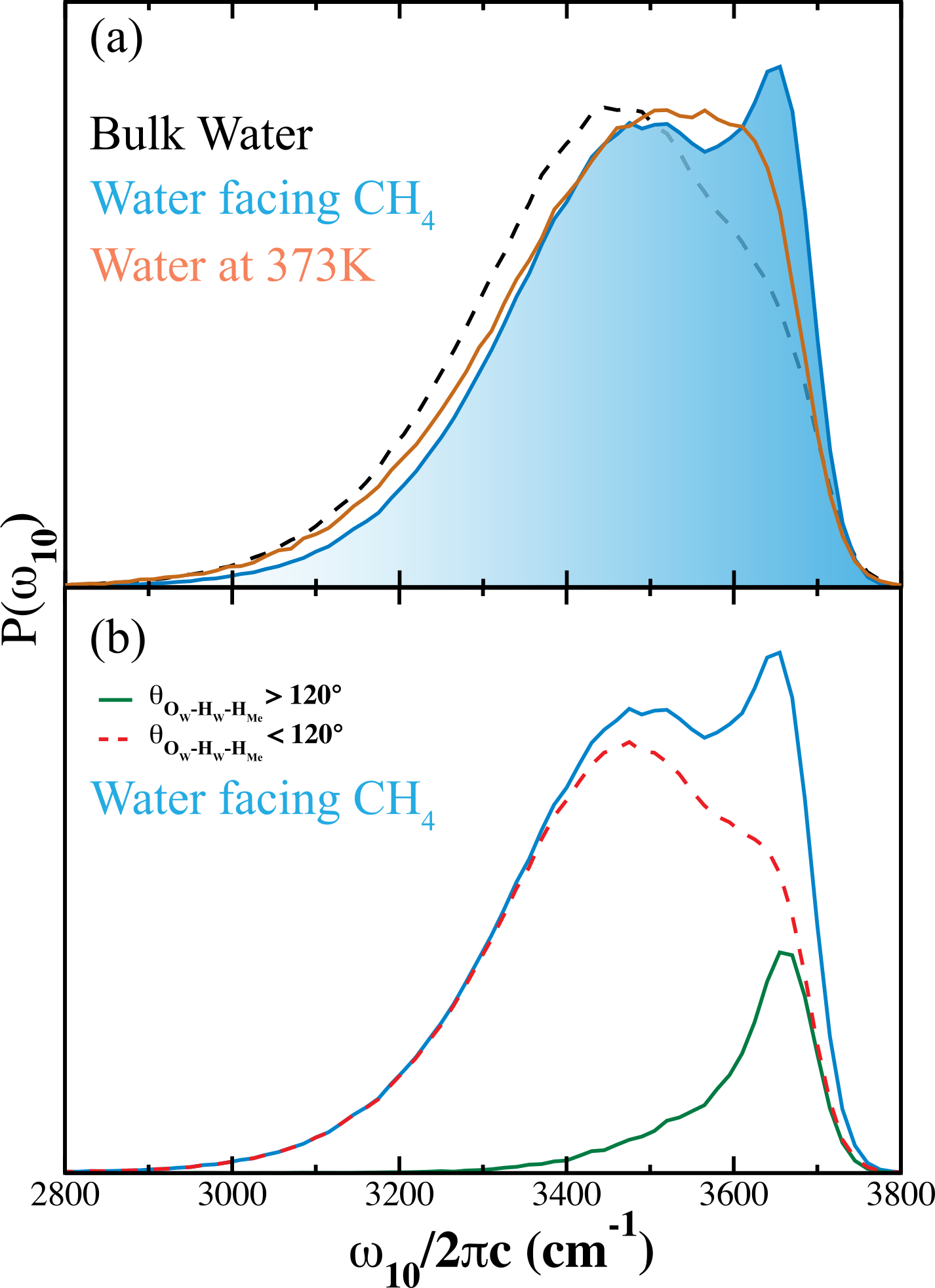}
\caption{\label{fig5} a) Comparison of stretching frequency $(\omega _{10})$ bulk water (black), closest water to methane (cyan) and water at 373 K (red). b) Stretching frequency $( \omega _{10} )$ of closest water to methane bimodal distribution divided based on angle.}
\end{figure}

To understand the observed blueshift in the O-H stretch frequency of the water molecule nearest to the methane moiety, we perform bulk water simulation at T=373 K. In Figure \ref{fig5}(a) we represent the O-H frequency distributions of bulk water, water molecule closest to methane and water at 373 K. We indeed observe that the frequency distribution in the case of nearest water molecule is much more similar to that of water at T=373 K (Figure \ref{fig5}(a)). This suggests that the nearest water molecule behaves like low-density water, in fact it is almost similar to the dangling water in case of air-water interface.\cite{Du1994} This observation supports the suggestion made by Stillinger as well.\cite{Stillinger1973} It is clear that the O-H bond pointing towards methane moiety is experiencing extremely low hydrogen bond environment which is originating because of the alteration of hydrogen bond network by the hydrophobic methane moiety. The presence of methane moiety makes the neighboring environment similar to that of surface water.\cite{Stillinger1973}
\par
However, this does not explain the two-ensemble nature of the distribution. To inspect that, we correlate the observed frequency with the $\angle \mathrm{O_WH_WH_{Me}}$ (Figure \ref{fig5}(b)). We find that the high frequency ensemble with larger population is coming from the configurations in which $\angle \mathrm{O_WH_WH_{Me}} > 120 \degree $ and the low frequency ensemble is originating from the configurations having $\angle \mathrm{O_WH_WH_{Me}} < 120 \degree $ (Figure \ref{fig5}(b)). Thus, the blueshift is increasing when the configurations are having more linear arrangement of the O-H bond of the nearest water molecule with the methane C—H bond. With increase in linearity, the propensity of interacting the O-H bond with the neighboring polar water molecules reduces, which results in higher blueshift in the observed transition frequency.\par
\begin{figure}
\includegraphics[width=.8\linewidth]{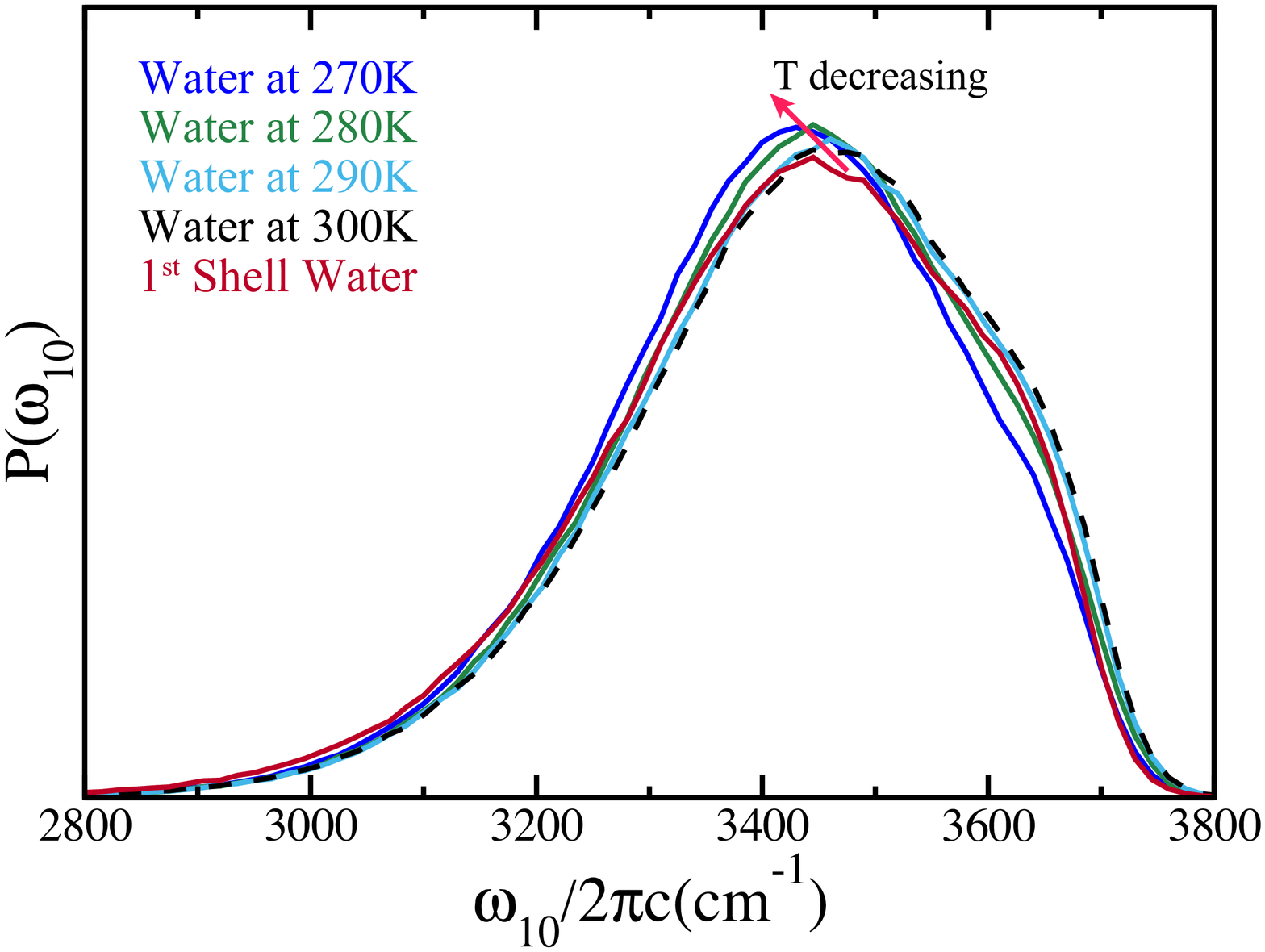}
\caption{\label{fig6}Comparison of frequency distributions of first solvation shell water and bulk water at T=300K, 290K, 280K and 270K.}
\end{figure}

We have already discussed that the $1^{st}$ solvation shell water molecules (except the water molecules which directly faces the methane moiety) of methane show redshifted O-H stretch frequency distribution (Figure \ref{fig4}). To explore the observed redshift in the O-H stretch frequency of the $1^{st}$ shell water, we perform bulk water simulation at T=290 K, 280 K and 270 K. In Figure \ref{fig6}, we represent the O-H stretch frequency distribution in case of first shell water, bulk water at T=300 K, 290 K, 280 K and 270 K. It is evident from the figure that the first shell water behaves like more structured low temperature water.
\par

\begin{table}[]
\centering
\caption{Average hydrogen bond numbers per water molecule in bulk water, $1^{st}$ shell, $2^{nd}$ shell and $3^{rd}$ shell water of methane molecule.}
\begin{tabular}{ll}
\hline
Contributing water ensemble	& Average $n_{HB}$\\
\hline
\hline
Bulk water	& 3.55\\
1st shell	& 3.51\\
2nd shell	& 3.54\\
3rd shell	& 3.54\\
\hline
\end{tabular}
\label{table4}
\end{table}

\begin{figure}
\includegraphics[width=0.8\linewidth]{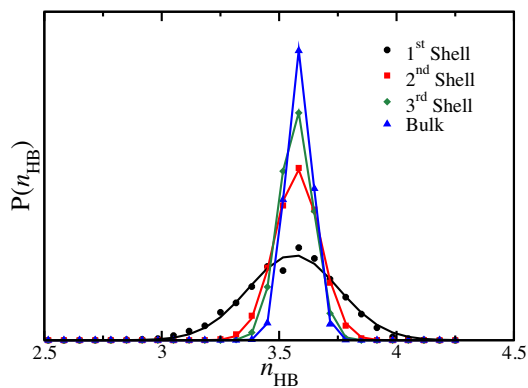}
\caption{\label{fig7}Comparison of hydrogen bond $(n_{HB})$ distribution in bulk water and different solvation shells of methane. All symbols represent the data and solid line represents the respective Gaussian fit.}
\end{figure}

In order to understand the above trends in the observed O-H stretch frequency, we further investigate the hydrogen bond distribution using standard geometrical criteria.\cite{Luzar1993} A water molecules is considered to form hydrogen bond with another water molecule if the inter-oxygen distance is less than 3.5 \text{\AA}, the hydrogen acceptor distance is less than 2.6 \text{\AA}, and H$_d$-O$_d$-O$_a$ angle is less than 30\degree, where the subscript ``$d$'' and ``$a$'' symbolize donor and acceptor respectively. The average hydrogen bond numbers in the different solvation shell water molecules are presented in Table \ref{table4}. We show the distribution of hydrogen bond number $(n_{HB})$ in each solvation shell along with the bulk water in Figure \ref{fig7}. It is evident from the figure is that the water molecules in the first solvation shell are having a much wider distribution spanning from $(n_{HB}) \sim 3.0$ to $(n_{HB}) \sim 4.0$. Although the average hydrogen bond number does not correlate with the observed frequency data, however the distribution makes a clear revelation of the microscopic picture. Thus, more structured water molecules are indeed present in the first solvation shell which eventually is reflected in the redshift of the observed O-H frequency. The extended low $(n_{HB})$ tail in the distribution of first shell water also supports the blueshift of the transition frequency in case of some of the first shell water molecules. On further inspection, we find that the later water ensemble mainly consist the water molecules, which are directly facing the methane moiety. In case of second and third shell water molecules, the hydrogen bond distributions are much narrower than that of the first shell water. These distributions are in fact slightly different than that of the bulk water. The observed frequency distributions in these two solvation shells also reflect that water in these two solvation shells behave almost similar like bulk water.
 
\section{CONCLUSIONS}
In this work, we develop a semiempirical quantum-classical spectroscopy map to investigate the effects of hydrophobic solute in the water solvation structure. We find a strong correlation with the single oscillator level O-H stretch frequency with a collective variable solvation energy. Our model predicts that the water hydrogen bond network gets significantly modified in presence of small hydrophobic molecule such as methane. A bimodal frequency distribution with a blue shifted population of transition frequency illustrates presence of more like low-density water molecules and dangling water molecules in the hydration shell of methane. We also find that there exist a strong correlation between the transition frequency of nearest water O-H bond pointing towards methane moiety and the orientation of that O-H bond with that of  C-H  bond of methane. Temperature dependent simulation depicts that the water molecules facing to the methane molecule behave similar to the high temperature water. The solvation shell-wise decomposition of the O-H stretch frequencies further established that there exists a significant increase in ordering of the first solvation water except the water molecules, which are directly facing the methane molecule. This is manifested in redshift of the observed transition frequencies. Temperature dependent simulation also depicts that the water molecules in the first solvation shell except the water molecules, which are facing to the methane molecule behave more like cold water.

\section{SUPPLEMENTARY MATERIAL}
See supplementary information in Appendix \ref{sinfo} for complete information regarding cluster selection, correlation between electric field and transition frequency, spectroscopic maps and empirical relationships for other vibrational transitions and construction of different solvation shells of methane hydration.

\section{Acknowledgements}
RB acknowledges IIT Tirupati for support through new faculty seed grant and also the computational support.


\clearpage
\appendix
\section{Supporting Infromation}\label{sinfo}
	Here we represent detail descriptions regarding cluster selection, correlation between electric field and transition frequency, spectroscopic maps and empirical relationships for other vibrational transitions and construction of different solvation shells of methane hydration.
\subsection{Cluster Selection}
We extract small clusters by identifying a central H atom that belongs to the water molecule closest to the methane molecule and include any molecules having its oxygen within a 7.0 \text{\AA} radius. In these clusters, average of $\sim 44$ water molecules are there within that cut-off distance, this number of water molecules are sufficient to produce bulk like environment surrounding the central water molecules. We follow similar approach for selecting the bulk water clusters. In methane water system we have found the minimum distance between methane carbon and water oxygen is 2.72 \text{\AA} and maximum distance is 17.17 \text{\AA}. We choose methane water clusters such that the selected clusters span over methane-water distance range of 2.85 \text{\AA} - 3.85 \text{\AA}. Furthermore, to investigate the role of orientation of water with respect to the methane  C-H  bond, we track the angle formed by H$_{\mathrm{Me}}$, H$_{\mathrm{W}}$ and O$_{\mathrm{W}}$, where H$_{\mathrm{Me}}$ signifies methane hydrogen, H$_{\mathrm{W}}$ and O$_{\mathrm{W}}$ are hydrogen and oxygen of water molecule closest to the methane moiety. Eventually, we segment the methane-water clusters into three ensembles based on the $\angle \mathrm {O_WH_WH_{Me}}$ angle values 150\degree -160\degree, 160\degree -170\degree and 170\degree -180\degree.
\renewcommand{\thetable}{S1}
\begin{table}[b]
	\centering
	\caption{Linear empirical relation between electric field and fundamental stretching frequency. We use fitting function $\omega_{10}=a_0+a_1E$ and the parameters and correlation coefficient are shown below.}
	\begin{tabular}{llll}
		\hline
		System			& $a_0$		& $a_1$		& Correlation Coefficient\\
		\hline
		\hline
		Bulk water		& 3775.89	& -8494.7	& 0.8409
		\\
		Methane water 	&	3711.1	& -343.27	& 0.1274
		\\
		\hline
	\end{tabular}
	\label{tables1}
\end{table}
\renewcommand{\thefigure}{S1}
\begin{figure}[htbp!]
	\centering
	\includegraphics[width=0.9\linewidth]{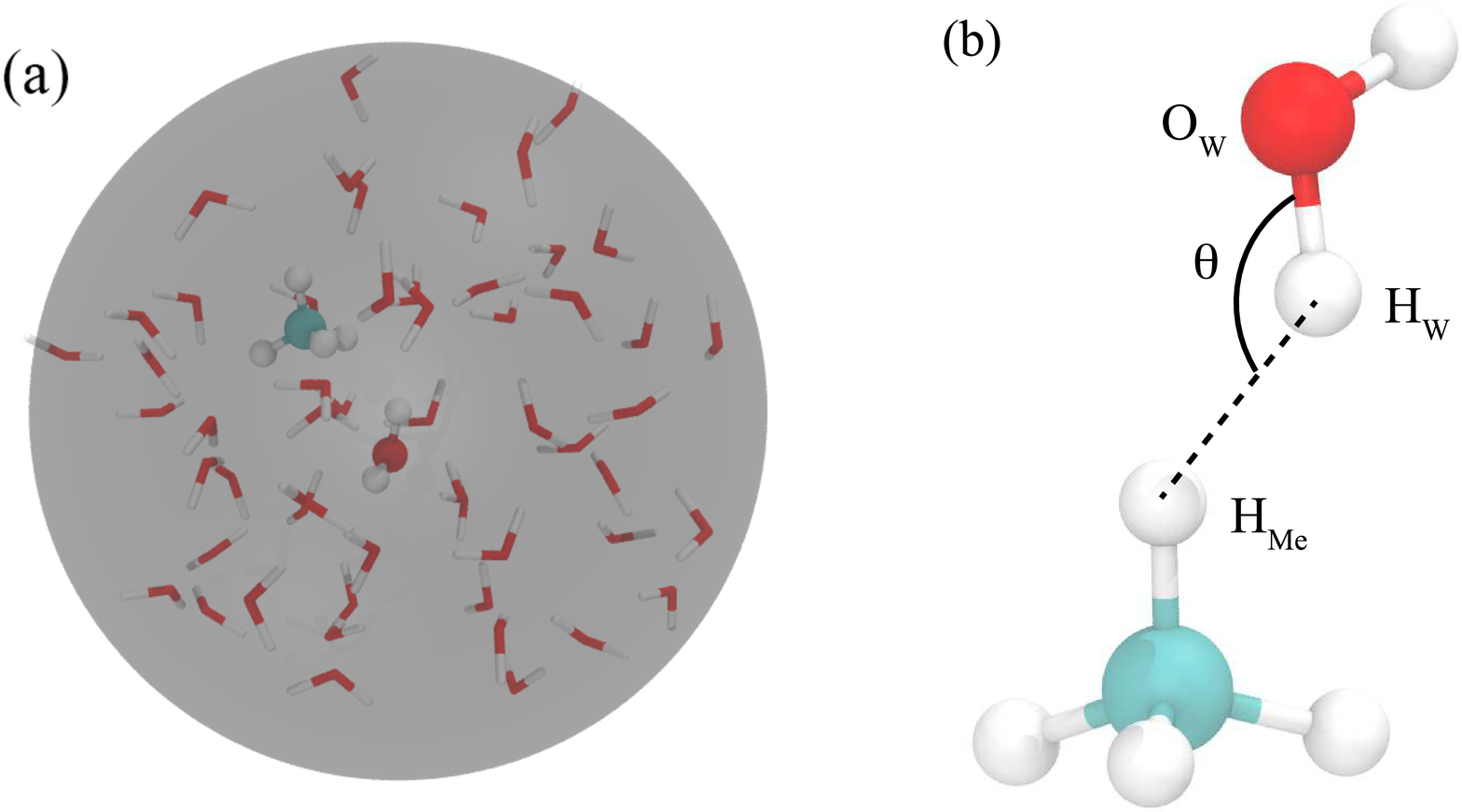}
	\caption{\label{figs1} (a) Representative snapshots of methane-water cluster as defined in the text. (b) Definition of the angle between methane hydrogen, water hydrogen and water oxygen.}
\end{figure}
\renewcommand{\thetable}{S2}
\begin{table*}[]
	\centering
	\caption{Empirical relations between the transition frequencies and the solvation energy obtained by fitting the quantum mechanical data as shown in Figure \ref{figs3}. We use the following fitting function: $\omega_{ij}=a_0+a_1\Delta E_{\mathrm{DFT}}+a_2\Delta E_{\mathrm{DFT}}^2+a_3\Delta E_{\mathrm{DFT}}^3+a_4\Delta E_{\mathrm{DFT}}^4$}
	\begin{tabular}{lllllll}
		\hline
		$\omega_{ij}$	& $a_0$	& $a_1$	& $a_2$	& $a_3$	& $a_4$	& Correlation coefficient\\
		\hline
		\hline
		$\omega_{21}$	& 541.83	& 0.3039	& $-2.8348 \times 10^{-6}$	& $-6.4550 \times 10^{-10}$	& $2.2335 \times 10^{-14}$	& 0.9990 \\
		$\omega_{20}$	& 1631.11	& 0.5663	& $-2.4002 \times 10^{-6}$	& $-1.5512 \times 10^{-9}$	& $5.2373 \times 10^{-14}$	& 0.9981 \\
		\hline
	\end{tabular}
	\label{tables2}
\end{table*}
\renewcommand{\thetable}{S3} 
\begin{table}[!ht]
	\centering
	\caption{Empirical relationship with and the dipole moment derivative scaled with respect to the gas phase dipole moment derivative with the form ${\mu^\prime}/{\mu^\prime_g}=a_0+a_1\Delta E_{\mathrm{DFT}}$}
	\begin{tabular}{llll}
		\hline
		$a_0$	& $a_1$	& Correlation coefficient\\
		\hline
		\hline
		1.8833	& $-8.4546 \times 10^{-5}$	& 0.8205\\
		\hline
	\end{tabular}
	\label{tables3}
\end{table}
\renewcommand{\thetable}{S4}
\begin{table*}[]
	\centering
	\caption{Empirical relation between the transition frequency and the transition dipole moment obtained by fitting the data as shown in Figure S5. The fitting function is $\mu_{ij}=a_0+a_1\omega_{ij}+a_2\omega_{ij}^2+a_3\omega_{ij}^3+a_4\omega_{ij}^4$}
	\begin{tabular}{lllllll}
		\hline
		$\mu_{ij}$	& $a_0$	& $a_1$	& $a_2$	& $a_3$	& $a_4$	& Correlation coefficient \\
		\hline
		\hline
		$\mu_{21}$ & 3.5600	& $-4.4836 \times 10^{-3}$	& $2.2114 \times 10^{-6}$	& $-4.8574 \times 10^{-10}$ &	$3.9424 \times 10^{-14}$	& 0.990 \\
		$\mu_{20}$ & 2.4638	& $-1.4258 \times 10^{-3}$	& $3.2558 \times 10^{-7}$ &	$-3.37681 \times 10^{-11}$	& $1.3189 \times 10^{-15}$	& 0.9999\\
		\hline
	\end{tabular}
	\label{tables4}
\end{table*}
\renewcommand{\thefigure}{S2}
\begin{figure}[]
	\centering
	\includegraphics[width=0.7\linewidth]{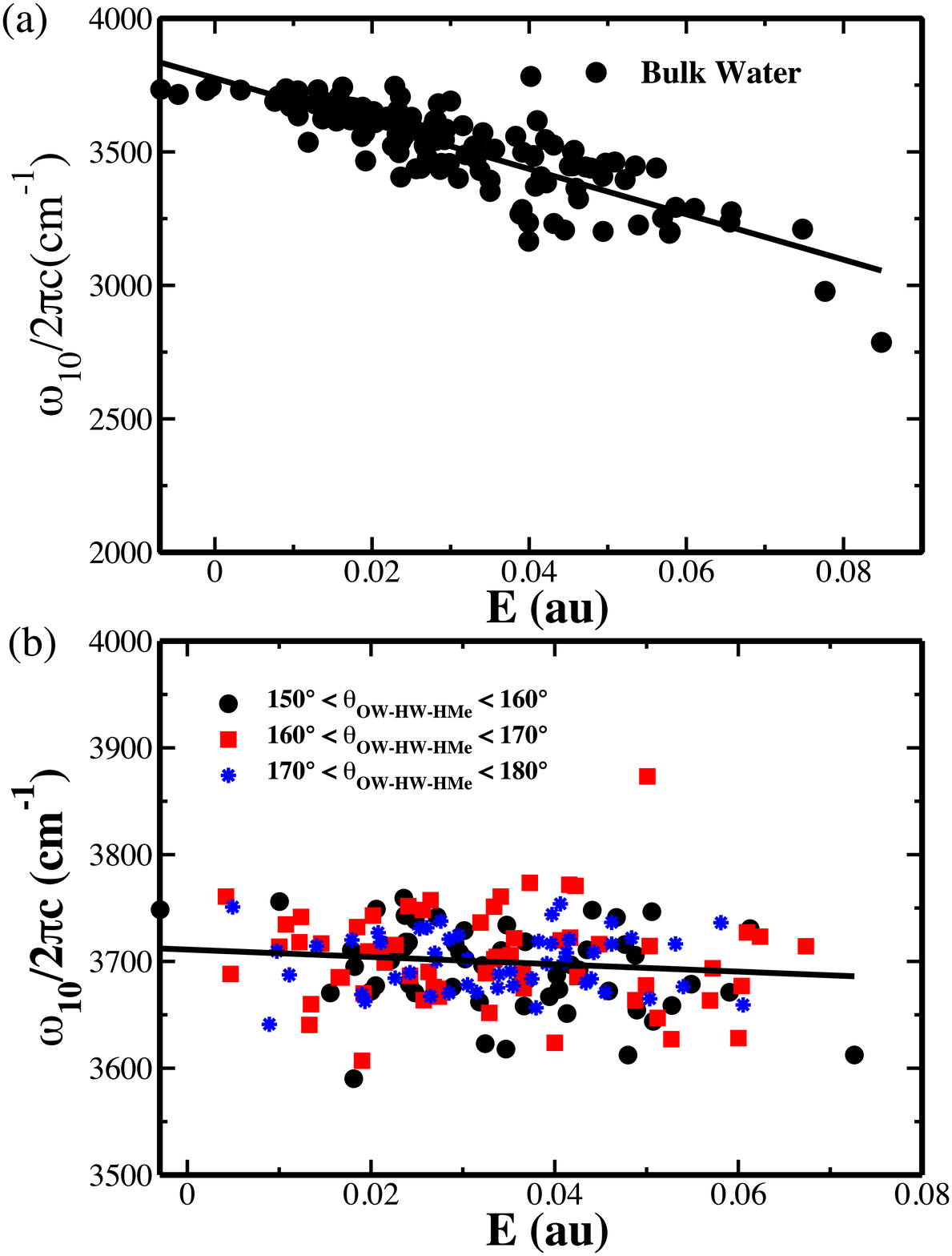}
	\caption{\label{figs2} Correlation between electric field $(E)$ projected along O-H bond and fundamental stretching frequency of bulk water (a) and methane water system (b).}
\end{figure}
\subsection{Correlation between Electric Field and Transition Frequency}
The electric field of the system projected along O-H bond calculated by following expression for all systems.\cite{Corcelli2005}
\renewcommand{\thefigure}{S3} 
\begin{figure}[H]
	\centering
	\includegraphics[width=0.7\linewidth]{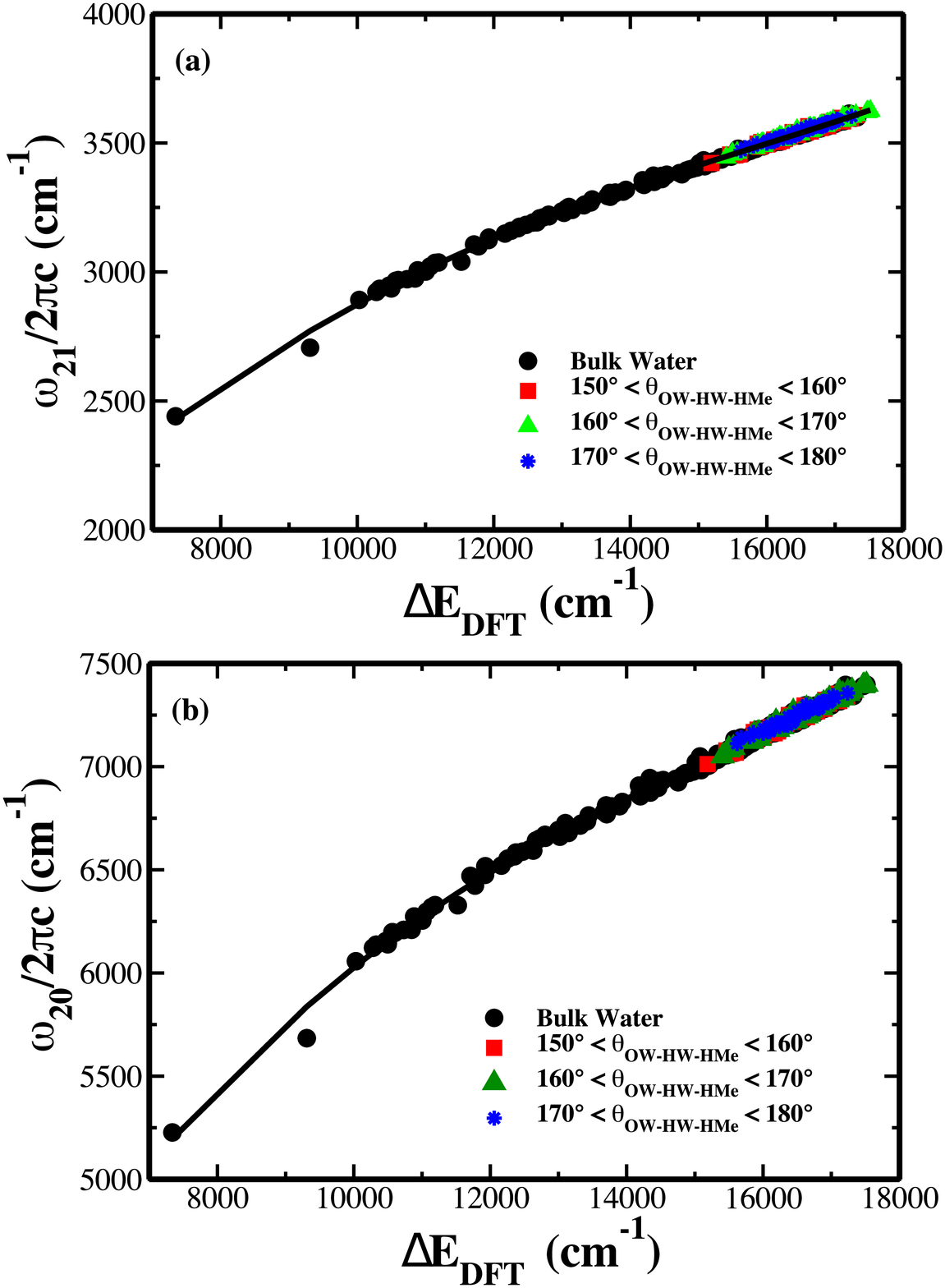}
	\caption{\label{figs3} Correlation of different transition frequencies with the solvation coordinate obtained from DFT calculation $(\Delta E_{\mathrm{DFT}})$. Solid black line represents the quadratic fit and the fitting parameters are provided in Table \ref{tables2}}
\end{figure}
\renewcommand{\theequation}{S1}
\begin{equation}
E = \hat{\textbf{u}}\cdot \sum_{i=1}^{mn} \frac{q_i\hat r_{iH}}{r_{iH}^2}
\end{equation}
Where $\hat{\textbf{u}}$ is the unit vector corresponding to the O-H bond of interest, $q_i$ is the charge of the $i^{th}$ site, $\hat r_{iH}$ is the distance between $i^{th}$ site and the H of center H$_2$O molecule, having $n$ molecules having $m$ charged atoms per molecules. Thus, $m$ is 3 for water molecules and 5 for methane molecules.
We use linear fitting function to examine the correlation and the fitting parameters are given in Table \ref{tables1}. It is clear from the figure that the bulk water data show significant correlations similar to the previous findings,\cite{Corcelli2005} however, the data for methane water system show very little correlation.
\subsection{Spectroscopy Maps}
\renewcommand{\thefigure}{S4} 
\begin{figure}[H]
	\centering
	\includegraphics[width=.8\linewidth]{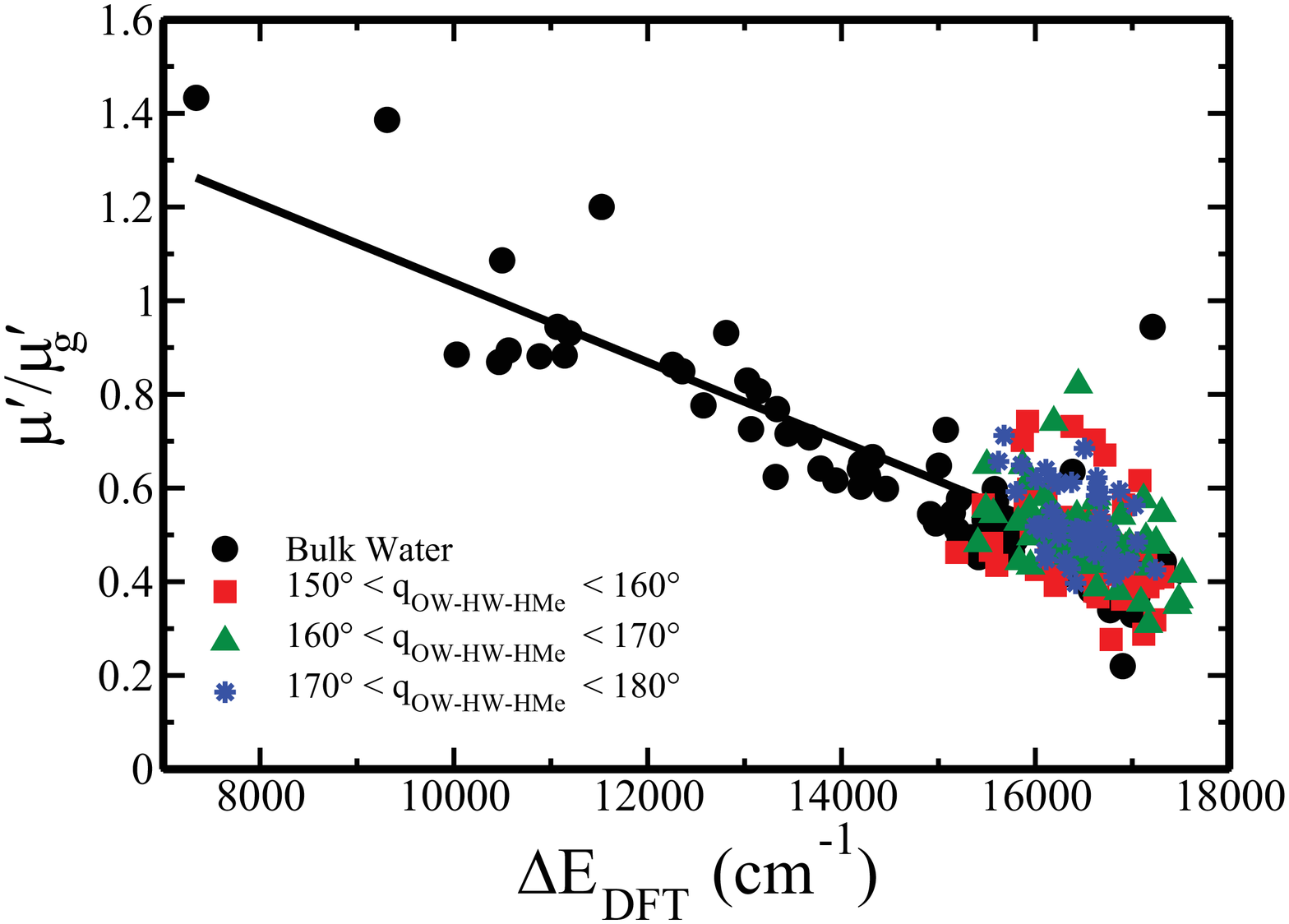}
	\caption{\label{figs4} Correlation of solvation energy calculated from DFT and dipole moment derivative scaled with respect to the gas phase dipole moment derivative. Black solid line represents the linear fit and the fitting parameters are given in Table \ref{tables3}.}
\end{figure}
We represent the correlation of $\omega_{21}$ and $\omega_{20}$ with the solvation energy obtained from the quantum mechanical calculation in Figure \ref{figs3}. Both the transition frequencies show a similar monotonic and highly correlated behavior. We fit the data with $4^{th}$ order polynomials. The fitting functions and the respective fitting parameters are provided in Table \ref{tables2}. We also investigated the effect of orientation of the tagged O-H bond with the methane molecule in a similar fashion as has been done for fundamental transition frequency. We find there exist minimal or almost no effect of the orientation of the O-H oscillator with respect to the methane moiety on the spectroscopy maps.\par
\renewcommand{\thefigure}{S5}
\begin{figure}[H]
	\centering
	\includegraphics[width=0.7\linewidth]{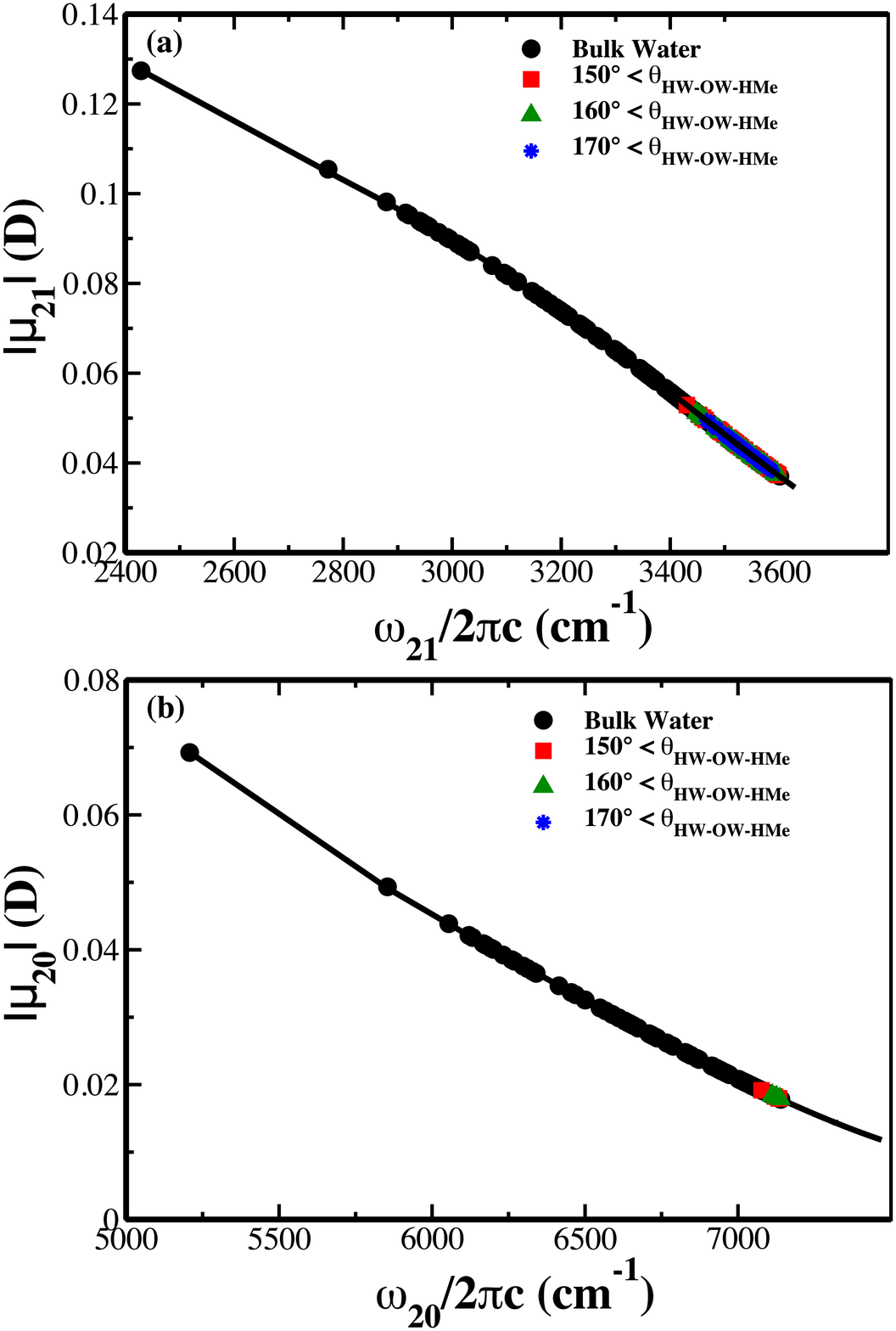}
	\caption{\label{figs5} Correlation between different transition dipole matrix elements with their rpective transition frequencies. The black line represents the quadratic fit to the data.}
\end{figure}
We obtain $\mu^\prime$ for each chosen cluster by calculating $\vec{\mu}_{nm} \cdot \hat{\textbf{u}}$ at five $r_{\mathrm{OH}}$  displacements separated by 0.01 \text{\AA} about $r_{eq}$, and then numerically differentiate with respect to $r_{\mathrm{OH}}$ . Finally, we construct the correlation of dipole moment derivative scaled with respect to the gas phase dipole moment derivative with the solvation energy calculated from DFT (Figure \ref{figs4}). The scaled dipole moment shows a linear correlation with correlation coefficient 0.8205 (Table \ref{tables3}).\par
We represent the correlations between different transition dipole moments with corrresponding transition frequencies in Figure \ref{figs5}. We observe a similar monotonic and highly correlated data as observed in the case of fundamental transition dipole moments and transition frequency. We use $4^{th}$ order polynomial fitting functions to get the empirical relations and the corresponding fitting parameters are shown in Table \ref{tables3})
\subsection{Construction of Solvation Shells}
We subdivide the water molecules surrounding methane molecule into three solvation shells based on their distances from the methane moiety. In Figure \ref{figs6}, we represent the radial distribution function of carbon methane and water oxygen. Based on the radial distribution data, we consider all the water molecules within 5.5 \text{\AA} distance from methane as first solvation shell, second solvation shell consists of water molecules residing within 5.5 \text{\AA} to 9 \text{\AA} and left over water molecules are considered as third solvation shell.
\renewcommand{\thefigure}{S6}
\begin{figure}[H]
	\centering
	\includegraphics[width=.8\linewidth]{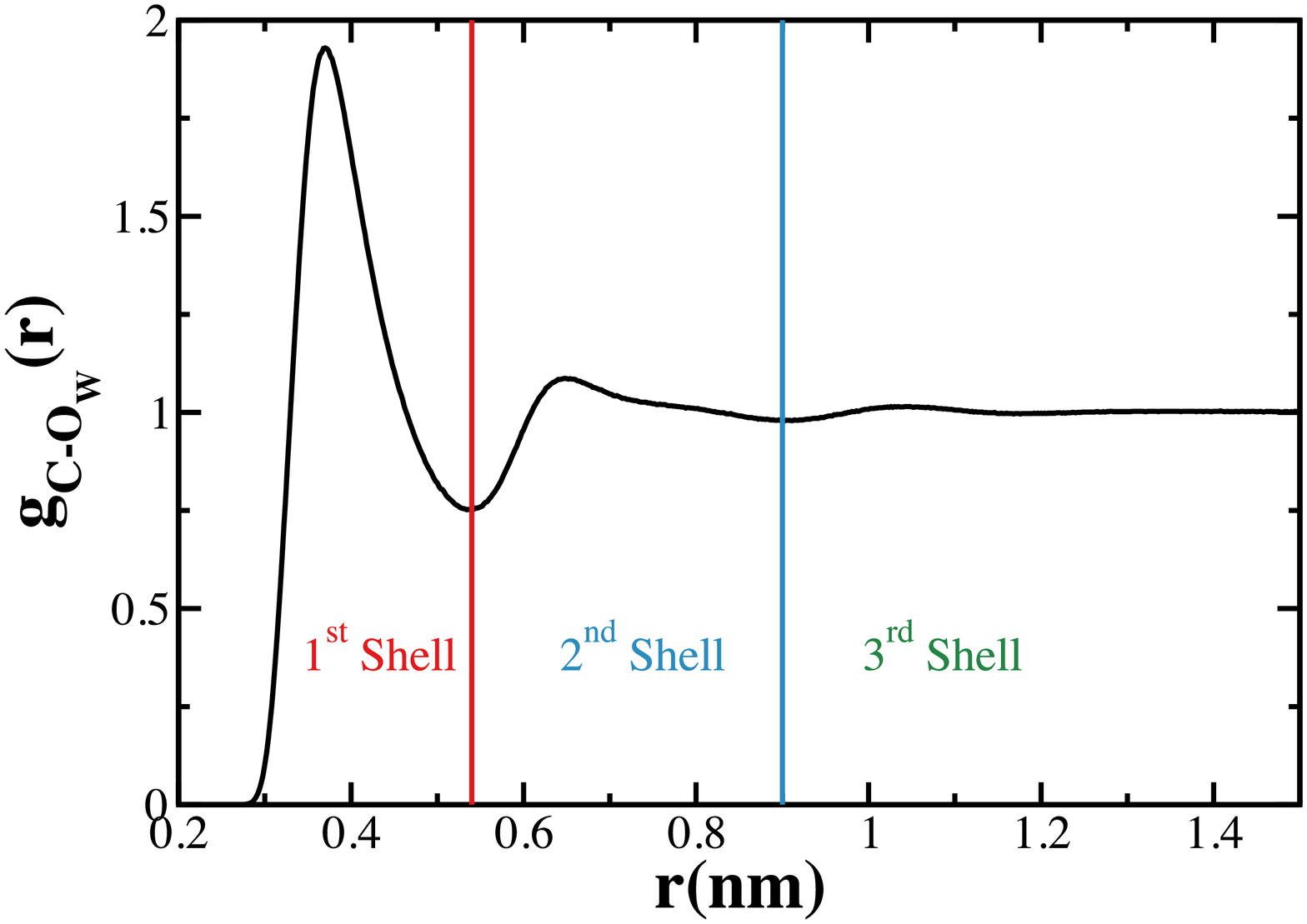}
	\caption{\label{figs6} Radial distribution function of carbon of methane and oxygen of water. Note that three ensemble divisions of solvation shells are highlighted in the figure.}
\end{figure}
\end{document}